\documentstyle[epsf]{laa}     

\begin{document}
   \thesaurus{07         
              (07.09.1;  
               03.01.1;  
               04.01.1)  
             }

\title{
Long-range (fractal) correlations in the LEDA database}
   \subtitle{}

   \author{H. Di Nella \inst{1,2}, M. Montuori \inst{3,4}, G. 
Paturel\inst{1}, 
            L. Pietronero \inst{4}, and F. Sylos Labini \inst{4}}
   \institute{
              CRAL - Observatoire de Lyon,
              69561 Saint-Genis Laval, France 
  \and
              Depart. astrophysics, Uni. New South Wales,
              2052 Sydney, NSW, Australia 
  \and
              Dip. di Fisica, Universit\'a della Calabria,
              Cosenza, Italy
  \and
       	      INFM Sezione Roma1,       
	      Dip. di Fisica, Universit\'a "La Sapienza",
	      P.le A. Moro, 2, 
              I-00185 Roma, Italy
}

   \date{Received -- -- --; accepted -- -- --}

   \maketitle

\newcommand{\ajk}{\alpha_{j,k}}
\newcommand{\be}{\begin{equation}}
\newcommand{\ee}{\end{equation}}
\newcommand{\bdis}{\begin{displaymath}}
\newcommand{\edis}{\end{displaymath}}
\newcommand{\eg}{\varepsilon}
\newcommand{\hmp}{h^{-1} Mpc}

\begin{abstract}
All the recent 
redshift surveys show highly irregular patterns 
of galaxies on scales of hundreds
of megaparsecs such as chains, walls and cells. One of the 
most powerful catalog of galaxies is represented by the LEDA 
database that contains more than 36,000 galaxies with redshift.
We study the correlation properties of such a sample
finding that galaxy distribution shows well
defined fractal nature up to R$_s \sim 150 \hmp $
with fractal dimension $D \approx 2$. 
We test the consistency 
of these results versus 
 the incompleteness in 
the sample. 
\end{abstract}

\section{Introduction }
 
Many thousands of galaxies have been catalogued across 
the whole sky and for some tens of thousands of them there are 
also available  redshift measurements. In the next few years 
the number of redshifts will increase very quickly by a factor of
ten or more. However, the present 3 dimensional data on the
distribution of galaxies already gives us the possibility to study and 
characterize 
quantitatively the visible matter distribution at least 
up to $150 \hmp$.
Several discoveries of large scale structure (LSS) were reported in 
recent
years. The first discovery of cell-like structure of the Universe
was reported by Einasto and co-workers (1983). Later, using the 
CfA1 
redshift survey, ( Huchra et al., 1983; 
De Lapparent et al., 1988) 
confirmed 
the existence of voids and discovered a filament in the Coma cluster 
region. 
Large Scale distribution of visible matter is characterized 
by having strong inhomogeneities of all scales so that the scale of 
the 
largest inhomogeneities is comparable with the extent of the 
surveys 
in which they are detected (Broadhurst et al., 1990; Vettolani et al., 
1994; 
Da Costa et al., 1994).

Anticipating the growth of the redshift industry of the last decade 
and 
pursuing the philosophy of the RC1 and RC2 catalogs, the Lyon-Meudon 
extragalactic database, 
LEDA, was created in 1983.
This database has grown
and now contains more than 100,000 galaxies with the most 
important
astrophysical parameters: names of galaxies, morphological 
description,
diameters, axis ratios, magnitudes in different colors, radial 
velocities,
21-cm line widths, central velocity dispersions, etc... 
In the 12 years since the inception of LEDA, 
more than 75,000 redshifts have been collected, for some 
36,000 galaxies.
From a Flamsteed's equal area projection made with galaxies of
LEDA the existence of
a very large structure was suspected (Bottinelli et al., 1986;
Paturel et al., 1988).
This structure was called hypergalactic because
it seems to connect several superclusters (Perseus-Pisces, Pavo-
Indus,
Centaurus and The Local Supercluster). 
Di Nella \& Paturel (1995) show that this structure corresponds to a 
privileged plane of the Local universe containing 45\% of the
galaxies where only 25\% would be expected if the galaxy distribution was
homogeneous.

Due to the existence of such complex LSS,
one of the present problems of experimental cosmology is to 
identify in 
the redshift data the scale of homogeneity, i.e. the scale over which 
a constant density of galaxies could be defined unambiguously.
In this paper, we look for 
{\it any} defined value of the scale of homogeneity
and whether the average density is a well defined
 quantity or not, up to $150 \hmp$. 
We show that these questions are connected with 
the intrinsic highly irregular nature of galaxy distribution.

There has been a great debate about the scale of homogeneity.
Coleman \& Pietronero (1992 - CP92)
 introduced a new statistical analysis that reconciles 
galaxy correlation with the existence of LSS. This is the appropriate 
correlation analysis for studying highly irregular distributions, 
but, of course, it can be successfully applied also in the case of
smooth (homogenous) distributions. 
CP92 showed that the interpretation of the standard analysis 
performed by 
the $\xi(r)$ two-points correlation function (Davis \& Peebles, 
1983)
is based on the {\it assumption that the sample under 
analysis is homogeneous}. 
In the case in which this
 basic hypothesis is not satisfied the $\xi(r)$ analysis 
gives misleading results because it gives information that 
are related to the sample size rather than to any real physical 
features.
CP92 found a fractal behaviour in the CfA1 redshift  
survey up to $\sim 20 \hmp$
 with a fractal dimension $D \approx 1.5$, 
and Sylos Labini et al. (1996 - SL96) found that 
in Perseus-Pisces surveys the fractal behaviour extends up to $130 \hmp$
with $D \approx 2$. 
Here we investigate the correlation properties
of the redshift samples available in LEDA database, 

\section{Description of the sample}

From LEDA, various magnitude limited redshift samples 
were extracted. They cover the whole sky.
We studied the incompleteness of LEDA in terms of rate of redshift 
measurements 
versus their apparent magnitude.
Figure 1 shows the number of 
galaxies with a known magnitude in LEDA 
in function of their  
apparent
magnitude (dashed line). Is also plotted the number of these 
galaxies having 
a measurement of redshift presently published (solid line). 
One can deduce that up to roughly B$_{t}$=14-14.5 the rate 
of measurement is 90\%.
The rate of measurement 
decreases as the limiting magnitude becomes fainter. A sample
selected in apparent magnitude with a limit at B$_{t}$=16 
will have a rate of 60\% and a sample selected with a limit of 
B$_{t}$=17 has 
a rate of measurement of 50\%.

\begin{figure}
\epsfxsize=8cm
\hbox{\epsfbox[39 294 534 663]{f1.ps}}
\caption{ To investigate  the completeness  of
the database in terms of rate of  measurements 
in a survey, we plot the number of galaxies
with a known magnitude (dashed line) and the
number of these galaxies having a measured redshift 
in LEDA (continue line). This plot shows the completeness for 
LEDA17
(half-sky). Up to $B_T \sim14.5$, 90\% of the known 
galaxies are measured. After this limit, the database begins to be 
incomplete. }
\label{fig1}
\end{figure}

In order to test the effect of incompleteness, we deliberately 
worked with 
complete ($m<14.5)$, partially incomplete ($m<16$) and incomplete 
($m<17$) samples. 
Even with our aim to subtract the geocentric bias 
involved in all catalogs surveying
or the southern or the northern equatorial hemisphere, we cut 
our all-sky samples in half-skies 
samples, in order to avoid the zones of avoidance due to the Milky 
Way ($b<-10^{\circ}$ and $b>10^{\circ}$).
No significant redshift data is presently available in these zones.
Hereafter we will call LEDA14.5, LEDA16 and LEDA17, the 
half-sky redshifts samples
 limited in apparent magnitude 
to $m_{lim}$=14.5,16 and 17,
respectively.
We have used heliocentric velocities and estimated the distance 
from the Hubble law.
From these samples we extract 
various subsamples limited in absolute magnitude and 
in volume (VL subsamples) as shown in Table 1.
\begin{table}
\caption{
The VL subsamples used.}
\label{subsamples}
\begin{tabular}{llll}
\hline
1/2 sky limited & VL & Limiting &  nb. of \\
in app. mag. & Mpc&  abs. mag. &  gal.    \\
LEDA14.5N & 80  & -20.07 & 563\\
LEDA14.5S & 80  & -20.07 & 605\\
LEDA16N   & 240 & -21.07 & 271\\
LEDA16S   & 240 & -21.07 & 465\\
LEDA17N   & 320 & -20.75 & 699\\
LEDA17S   & 320 & -20.75 & 1031\\
\hline
\end{tabular}
\end{table}
The stress that the magnitude of the galaxies registred in the 
LEDA database came 
from many different references.
We refer the reader to 
 Paturel et al. (1994) for a discussion of methods to 
 reduce all the different magnitude systems to the $B_t$ 
system.  The major part of the galaxies have a mean error
 less that 0.5 magnitude.

\section{Correlation function analysis}

The average density for 
 a sample of
 radius $\:R_{s}$ which contains a portion 
of the fractal structure is (CP92)
\be
\label{l5}
<n> =\frac{ N(R_{s})}{V(R_{s})} = \frac{3}{4\pi } B R_{s}^{-(3-D)}
\ee
From Eq.\ref{l5} we see that the average density 
is not a meaningful concept in a fractal 
because it depends explicitly on the sample 
size $\:R_{s}$.  
We can define the 
{\it conditional density} from an 
occupied point as (CP92):
\be
\label{l6}
\Gamma (r)= S(r)^{-1}\frac{ dN(r)}{dr} = \frac{D}{4\pi } B r ^{-(3-D)} 
=
\frac{<n(\vec{x}+\vec{r})n(\vec{x})>}{<n>}
\ee
where $\:S(r)$ is the area of  a spherical shell of radius $\:r$
($\gamma=3-D$ is the codimension).
The conditional average density, 
as given by Eq.\ref{l6}, is well defined in terms of 
its exponent, the fractal dimension. The 
amplitude of this function ($B$) is essentially  
related to the 
lower cut-offs of the fractal structure (CP92).
We limit our analysis to  the depth 
$R_s$ corresponding to  the radius of the 
maximum sphere fully contained in the
sample volume, in such a way we do not make use of any weighting 
schemes
and we do not introduce any {\it a priori} hypothesis on the nature of 
galaxy 
distribution.

 We adopted the following procedure to be sure 
that the possible incompleteness in the samples does not 
affect the correlation properties.
The analysis of a sample of galaxies 
in the same region of CfA1 is done 
with the same apparent 
magnitude limit $m_{lim}=14.5$. 
In this way we reproduce the 
results of CP92 as shown in Fig.2A:
the conditional density scales as $\Gamma(r) \sim r^{-\gamma}$ up to the sample
limit of $R_s \approx 20 \hmp$  with $\gamma=3-D \sim 1$ 
(Eq.\ref{l6}).
Then we increased the solid angle of the sample.
In this way we may increase step by step the 
actual radius $R_s$ up to which we compute the 
correlation function $\Gamma(r)$, 
so that we can control any bias 
or selection effects that can be introduced in 
the enlarged sample and may affect the correlation properties.
We find that the scaling region of $\Gamma(r)$ grows 
with the survey volume, and there is not any significant 
change neither of the slope nor of the amplitude of $\Gamma(r)$. 
When we take the maximum possible sphere fully contained 
in LEDA14.5,
that is the whole northern galactic hemisphere
above the milky way LEDA14.5N and LEDA14.5S for the southern
galactic hemisphere\footnote{ LEDA14.5N is defined by the range in 
galactic coordinates : $l[0,360]$ 
and $b[10,90]$. LEDA14.5S is defined by: $l[0,360]$ and $b[-90,-10]$.},
we find that 
$\Gamma(r)$ continues to scale up to $\sim 25 \hmp$
in the north and up to $\sim$ 30 $\hmp$ in the southern galactic hemisphere 
(Fig.2A).

\begin{figure}
\unitlength=1cm
\begin{picture}(9,14.3)
\put (0,9.3){\epsfxsize=8.5cm \epsfysize=4.5cm 
\hbox{\epsfbox[17 46 777 551 ]{f2a.ps}}}
\put (0,4.65) 
{\epsfxsize=8.5cm \epsfysize=4.5cm \hbox{\epsfbox[17 46 777 551 ]{f2b.ps}}}
\put (0,0) {\epsfxsize=8.5cm \epsfysize=4.5cm 
\hbox{\epsfbox[17 46 777 551 ]{f2c.ps}}}
\end{picture}
\caption{ 
{\it (A)} The correlation function $\Gamma(r)$ for the volume
limited at $80 \hmp$ of LEDA14.5:
the squares refer to the northern galactic hemisphere and the triangles to 
the southern 
one. 
{\it (B)} The same of (a) but for the volume limited samples
at $240 \hmp$ of LEDA16 north (squares) and south 
(triangles). 
{\it (C)} The same of (a) but for the volume limited samples
at $320 \hmp$ of LEDA17 north (squares) and south 
(triangles). 
The reference line has a slope $\gamma = -0.9$
for all the figures. The different amplitude of $\Gamma(r)$ 
for north and south samples in (b) and (c), is due to a different sampling rate
(Table 1). The different amplitude between  (a), (b) and (c) is 
due to the different luminosity selection of the VL samples
(Eq.3).
}
\label{fig2c}
\end{figure}
We are quite confident that the correlation properties
detected are genuine features of the sample because the possible
incompleteness or selection effects can only make worse the 
correlation properties of the sample but cannot create 
such correlations in full agreement with the properties 
found in the samples in which we 
know there is no such incompleteness.

The amplitude of $\Gamma(r)$ is related on the lower cut-offs of
the distribution and it  is simply connected to the prefactor
of the average density.
 To normalize the conditional density in different VL samples
(defined by the absolute magnitude limit $M_{lim}$)
 we divide it by 
\be
\label{x1}
\Phi(M_{lim}) = \int_{-\infty}^{M_{lim}} \phi(M) dM
\ee
where $\phi(M)$ is the Schechter luminosity function.
As parameter of $\phi(M)$ we adopt $\alpha=-1.1$ and $M_*=-
19.7$
(Da Costa et al., 1994).
We find that there is a nice match of the 
amplitudes and slopes. This 
implies that we are computing the right galaxy number density in 
each
subsample, and that this average density does not depend (or it
 depends very weakly) on the absolute magnitude of galaxies. 

Once that we have checked that the whole sky survey 
LEDA14.5 correctly reproduces global properties we can study the 
samples LEDA16 and LEDA17, that we
know to be affected by incompleteness (see {\it section 2}).
The problem is to understand if the incompleteness 
in the redshift collection  is strong enough to
destroy the correlation that we have found with LEDA14.5, or not.
This crucial point is strictly related to
  the following one: if, from a fractal 
distribution, we eliminate some points
randomly, and if the sample is statistically consistent,
we may reconstruct the right correlation properties.
On the contrary, if the 
point-cutting procedure is not random but  
 is systematic in some particular regions of the sky,
 the correlation properties can be seriously 
affected, and we should not recover the 
genuine properties of the sample.
As a test, we will adopt the following one: we compare the 
correlation 
properties of LEDA16 and LEDA17 with those of LEDA14.5, and 
in particular the fractal dimension and the amplitude of 
$\Gamma(r)$.
If the samples are seriously affected by
 selection effects the correlation 
properties will be changed and made noisy 
or completely destroyed; hence if  the sample 
at larger distances will show a cut-off towards homogeneity we 
should 
do a careful analysis of such a tendency 
because it should be due
to selection effects. On the contrary if the samples LEDA16 
and LEDA17 will show the same 
correlation properties than LEDA14.5, we will 
be quite confident that these 
 are the genuine physical features of the sample 
and are not an artifact due to selection effects and 
incompleteness in the data.

In Fig.2B we show the $\Gamma(r)$ for the various VL samples
of LEDA16. 
The agreement 
between  $\Gamma(r)$ of LEDA14.5 and LEDA16 (north and south) is very 
satisfactory. 
In Fig.2C we show the same behaviour 
for the VL samples of LEDA17. Also in this case the agreement is  
quite good.
Of course, the amplitude in the VL samples of 
LEDA16 and LEDA17 is lower than in the VL of LEDA14.5 (with the 
same 
absolute magnitude limit) because these two catalogs represent
a random sample of the whole galaxy distribution 
up to $m=16$ and $m=17$. As explained in before 
for LEDA16 the ratio of galaxies with a known magnitude
having a measured redshift is  $\sim 0.6$
while for LEDA17 this ratio is $\sim 0.5$.
The values of the amplitude of $\Gamma(r)$,
are consistent with these ratios (Di Nella et al., 1995).
Moreover the difference in amplitude in Fig.2B and Fig.2C
is due to a different sampling rate in the northern and in the 
southern
hemispheres.

We stress that $\Gamma(r)$ is an average statistical
quantity that measures a global properties of the 
distribution.
It is possible to check that eventual random error in the
determination of the apparent magnitude do not change substantially 
the fractal dimension (see Di Nella et al., 1995).
Moreover as $\Gamma(r)$ is computed performing an average over all 
the points of the sample, it is weakly affected 
by an eventual incompleteness of the sample with distance.

From Fig.2 we can conclude that LEDA17 shows long-range 
(fractal) correlation  up to the limiting depth of 
$R_s \sim 150 \hmp$
 with fractal dimension $D \approx 1.9$. Moreover we can 
conclude that the incompleteness effects are randomly 
distributed in the sky and do not affect crucially the 
correlation properties.
For a more complete discussion we refer
the reader to Di Nella et al., (1995) and
Di Nella \& Sylos Labini (1995).

With the aim of clarifying some basic aspects of 
the standard analysis (Davis \& Peebles, 1983),
 we have computed the two point
correlation function $\xi(r)$ for some VL samples 
of LEDA14.5, LEDA 16 and LEDA17.
$\xi(r)$ is  usually defined as:
\be
\label{s1}
\xi(r)= \frac{<n(\vec{x})n(\vec{r}+\vec{x})>}{<n>^2}-1
 = \left( \frac{3-\gamma}{3} \right) \left(\frac{r}{R_s}\right)^{-
\gamma} -1
\ee
Where the last equality holds for fractal distributions (CP92),
and $\gamma$=3-D.
As we stressed previously the $\xi(r)$ analysis 
gives misleading results in the case of fractal distribution,
and in any case it is not adequate to study fractal versus 
homogeneity properties.
As a fact, from Eq.4,  $\xi(r)$ is not a power law in the case of fractal
distribution and its amplitude depends on the sample size $R_s$.
The so-called correlation length $r_0$ defined as $\xi(r_0)=1$
is a linear function of the sample size:
$r_0=((3-\gamma)/6)^{\frac{1}{\gamma}} R_s
$.
\begin{figure}
\epsfxsize=8.5cm
\epsfysize=5cm
\hbox{\epsfbox[14 12 773 570]{f3.ps}}
\caption{ The so-called correlation length ($\xi(r_0)=1$) as 
a function of the sample size $R_s$ in LEDA17.
The fitting line has a slope $1$, compatible with 
the fractal behavior.
We find this same behaviour in the other samples of 
LEDA14.5 and LEDA16.}
\label{fig3}
\end{figure}
We have studied the behaviour of $r_0$
with sample size for LEDA14.5, LEDA16 
and LEDA17. 
The result (Fig.3) is in perfect agreement with the previous
prediction (Eq.4) as $r_0$ linearly scales with $R_s$, 
and thus with the fractal
behaviour of the galaxy distribution in the sample.

We can conclude that this analysis is in perfect agreement 
with the $\Gamma(r)$ analysis described in the previous section.
Several authors (Davis et al., 1988; Park et al., 1994) 
have tried to explain the scaling of $r_0$
with the sample size as due to the luminosity 
segregation phenomenon. It clearly appears from 
Fig.3. that this is not the case, 
as a "correlation length" of $r_0 \approx 45 \hmp$ is
clearly incompatible with such a phenomenon, 
and that the linear
scaling is fully compatible with the fractal nature 
of galaxy distribution (see also CP92 and SL96).

In conclusion, the long-range correlation present in this sample
are quite in agreement with the results of the analysis of 
CfA1 (CP92), 
Perseus-Pisces
(SL96, Guzzo et al., 1992), 
CfA2 (Park et al., 1994, Sylos Labini \& Amendola, 1995)
and finally in the deep ESP survey (Pietronero \& Sylos Labini, 
1995),
and confirm the evidence of fractal behaviour up to the largest
observed scales. 

These results seem to be in contrast with the 
analysis of the IRAS samples (Fisher et al., 1994) and the 
rescaling of the amplitude 
of the angular correlation function (Maddox et al., 1990)
that seem to be compatible with an homogenous distribution of (visible) matter. 
However it is possible to show (Sylos Labini et al., 1996) that
both these tests are strongly biased by spurious finite size effects: 
in the case of IRAS samples there are too few points to recover the genuine
features of the distribution, while in the case of the angular corelation
function, one determines its amplitude {\it without} performing any average
over different observers. In both these cases the apparent homogeneity 
can be shown to be related to finite size and selection effects.

\section*{Acknowledgements}We are grateful to those who have 
managed
the LEDA extragalactic database over the last 12 years:
N. Durand, P. Fouqu\'e, A.M. Garcia, R.Garnier, M. Loulergue,
M.C. Marthinet and C. Petit.

\end{document}